\documentclass[aps,pre,superscriptaddress, longbibliography,twocolumn,floatfix,noeprint]{revtex4-2}

\usepackage{amsmath}
\usepackage{amssymb}
\usepackage{amsfonts}
\usepackage{graphicx}
\usepackage{dcolumn}
\usepackage{bm}
\usepackage{latexsym}
\usepackage{slashed}
\usepackage{xcolor}
\usepackage[colorlinks=true,linkcolor=blue,citecolor=blue]{hyperref}
\usepackage{array}

\begin{document}

\title{Modeling ultrarelativistic streaming plasma instabilities under the quasistatic approximation}

\author{P.~San Miguel Claveria}
\affiliation{GAP/Instituto de Plasmas e Fusão Nuclear, Instituto Superior Técnico, Universidade de Lisboa, Lisbon, 1049-001, Portugal}

\author{Q.~Labro}
\affiliation{GAP/Instituto de Plasmas e Fusão Nuclear, Instituto Superior Técnico, Universidade de Lisboa, Lisbon, 1049-001, Portugal}

\author{X.~Davoine}
\affiliation{CEA, DAM, DIF, F-91297 Arpajon, France}
\affiliation{Universit\'{e} Paris-Saclay, CEA, LMCE, F-91680 Bruy\`{e}res-le-Ch\^{a}tel, France}

\author{A.~Matheron}
\altaffiliation[Present address: ]{Helmholtz-Institut Jena, Fr\"obelstieg 3, 07743 Jena, Germany}
\affiliation{Laboratoire d’Optique Appliquée (LOA), CNRS, École polytechnique, ENSTA, Institut Polytechnique de Paris, Palaiseau, France}

\author{M.~Tamburini}
\affiliation{Max-Planck-Institut f\"ur Kernphysik, Saupfercheckweg 1, D-69117 Heidelberg, Germany}

\author{S.~Corde}
\affiliation{Laboratoire d’Optique Appliquée (LOA), CNRS, École polytechnique, ENSTA, Institut Polytechnique de Paris, Palaiseau, France}

\author{L.~Gremillet}
\affiliation{CEA, DAM, DIF, F-91297 Arpajon, France}
\affiliation{Universit\'{e} Paris-Saclay, CEA, LMCE, F-91680 Bruy\`{e}res-le-Ch\^{a}tel, France}

\author{F.~Fiuza}
\email[Corresponding authors:\\ ]{pablo.san.miguel.claveria@tecnico.ulisboa.pt\\ sebastien.corde@polytechnique.edu\\
laurent.gremillet@cea.fr\\ frederico.fiuza@tecnico.ulisboa.pt }
\affiliation{GAP/Instituto de Plasmas e Fusão Nuclear, Instituto Superior Técnico, Universidade de Lisboa, Lisbon, 1049-001, Portugal}

\date{\today}

\begin{abstract}
Plasma streaming instabilities excited by relativistic charged particle beams play a pivotal role in astrophysical and laboratory environments. Their numerical study, however, is challenged by the disparity in spatiotemporal scales between the background plasma and beam particles, which can differ by several orders of magnitude for tenuous, ultrarelativistic beams. 
Here, we exploit the quasistatic approximation (QSA) to develop a new theoretical framework capable of capturing the full unstable spectrum in the spatiotemporal regime relevant for beams that continuously encounter unperturbed plasma at their leading edge. Within this linear, fully electromagnetic model, we uncover a previously unreported spatiotemporal evolution of the current filamentation instability and elucidate its interplay with the oblique two-stream instability, predicting the dominance of filamentation in the vicinity of the beam front. The good agreement between theory, kinetic particle-in-cell (PIC) simulations, and QSA-based PIC simulations validates the robustness of the approach. 
By pushing QSA-based PIC simulations to extremely dilute electron-positron beams, such as those found in blazar jets, we demonstrate their unique ability to capture the rich nonlinear dynamics of streaming instabilities in parameter regimes previously inaccessible to kinetic simulations.
\end{abstract}

\maketitle

\textit{Introduction.---}
Streams of relativistic charged particles propagating in collisionless plasmas are susceptible to various instabilities that can amplify electrostatic and electromagnetic fluctuations to the point of governing the evolution of the system~\cite{Sudan_Handbook_1984, Bret_POP_2010}. These collective phenomena play a fundamental role in energetic plasma environments, spanning both astrophysical and laboratory scenarios.

In astrophysics, streaming instabilities are thought to mediate magnetic field amplification, particle acceleration and radiation emission in collisionless shock waves, like those arising in supernova remnants and gamma-ray bursts ~\cite{Spitkovsky_APJ_2008a, Spitkovsky_APJ_2008b, Sironi_APJ_2009, Martins_APJ_2009, Lemoine_PRL_2019, Meszaros_NRP_2019, Peterson_PRL_2021, *Peterson_APJL_2022, Groselj_APJ_2022, *Groselj_APJL_2024, Vanthieghem_PRL_2024}. They are expected to regulate the transport of cosmic rays ~\cite{Marcowith_POP_2021} and may also affect the propagation of electron-positron pair beams from TeV blazars~\cite{Neronov_Science_2010, Broderick_APJ_2012, Sironi_APJ_2014}.

In the laboratory, they are critical for energy transport in ultraintense laser-plasma interactions \cite{Silva_POP_2002, Sentoku_PRL_2003, Adam_PRL_2006, Robinson_NF_2014, Schoenwaelder_NC_2026}, plasma-based accelerator concepts~\cite{Huntington_PRL_2011, Pukhov_PRL_2011}, or advanced $\gamma$-ray sources~\cite{Benedetti_NP_2018, Gong_PRL_2023}. This significance has driven efforts to investigate their fundamental properties in well-controlled accelerator experiments, using $\sim 60\,\rm MeV$ electron beams~\cite{Allen_PRL_2012} or $\sim 400\,\rm GeV$ proton beams~\cite{Verra_PRE_2024}.  Ongoing work includes studies at SLAC's FACET-II facility using $10\,\rm GeV$ electron beams~\cite{Yakimenko_PRAB_2019} and at CERN's HiRadMat facility using quasineutral $\sim 100\,\rm MeV$ electron-positron beams~\cite{Arrowsmith_NC_2024, Arrowsmith_PNAS_2025}.

Alongside kinetic theory, numerical particle-in-cell (PIC) simulations have been instrumental in identifying the oblique-two-stream (OTSI) and current filamentation (CFI) instabilities as the key players during the linear and early nonlinear stages of initially unmagnetized, relativistic beam-plasma interactions~\cite{Bret_POP_2010, Sironi_APJ_2014}. Yet modeling these instabilities remains very challenging when a large dynamical timescale separation exists between the beam and plasma particles, a situation typical of high-energy accelerator experiments and astrophysical systems. A realistic treatment must also account for the beam continuously encountering unperturbed plasma as it propagates. This inherently introduces spatiotemporal effects~\cite{Bers_Handbook_1983, Shukla_NJP_2020} missed by standard temporal theories~\cite{Bret_POP_2010}. 
Existing spatiotemporal models in the relativistic regime~\cite{Pathak_NJP_2015, San_Miguel_Claveria_PRR_2022} typically assume unstable modes are either fully electrostatic (OTSI) or inductive (CFI), failing to describe their mixed electromagnetic nature or competition in a unified setting. Moreover, these models hinge on the slowly-varying envelope approximation (SVEA), limiting their applicability to beams with longitudinal scales far exceeding the background plasma's skin depth ($k_p^{-1}$).

In this Letter, we leverage the quasistatic approximation (QSA) to develop a spatiotemporal theory of ultrarelativistic streaming instabilities that captures the full electromagnetic spectrum. By avoiding the SVEA, we track the onset of unstable modes in the vicinity of the beam front for the first time. This illuminates the interplay between CFI and OTSI under conditions where temporal models predict dominance of the latter. While we confirm this prediction deep into the beam---albeit with the spatiotemporal variant of OTSI dominant there---,we find that CFI, in its spatiotemporal form, prevails within a few $k_p^{-1}$ of the front, precisely where the SVEA breaks down. This contradicts the previous prediction that spatiotemporal CFI is negligible in the ultrarelativistic regime~\cite{Pathak_NJP_2015}.
We then provide a one-to-one comparison between this analytical model, PIC simulations, and the results of a new QSA-based PIC (QS-PIC) code~\cite{Labro_CPC_2026} which allows for dramatic computational speedup. Finally, we apply this simulation framework to a previously inaccessible blazar-relevant, ultra-dilute beam-plasma system.

\textit{Quasistatic model.---}
We consider a relativistic charged particle beam streaming through a charge-neutral plasma composed of immobile ions and initially stationary electrons with number density $n_p$. The beam may include several species (subscript $s$), each with charge $Z_s e$, mass $M_s m_e$ and number density $n_{bs}$ such that $n_{bs}/n_p \equiv \alpha_s \ll 1$. All beam species propagate at the same velocity $\mathbf{v_b} = \beta_b c \mathbf{\hat{x}}$ with $\beta_b \simeq 1$, corresponding to a Lorentz factor $\gamma_b = 1/\sqrt{1-\beta_b^2} \gg 1$. Here, $e$ denotes the elementary charge, $m_e$ the electron mass, and $c$ the speed of light. Our model adopts the cold-fluid approximation and is restricted to the $x$--$y$ plane. Equilibrium and first-order perturbed quantities are denoted with superscripts $^{(0)}$ and $^{(1)}$, respectively.

The QSA decouples the dynamics of the two interacting populations: the background plasma electrons, evolving on the plasma frequency timescale ($\omega_p = \sqrt{n_p e^2/(m_e \epsilon_0)}$, where $\epsilon_0$ is the vacuum permittivity), and the beam particles, evolving on their much slower (relativistic) plasma frequency timescale ($\omega_{bs} = Z_s \sqrt{\alpha_s/(\gamma_b M_s)}\omega_p \ll \omega_p$). The QSA is implemented by changing to comoving coordinates $(\xi,\tau) = (ct-x,t)$, and assuming that the plasma and field profiles vary much faster with $\xi$ than with $\tau$~\cite{Chen_IEEE_1987}. While the QSA is standard in the plasma-wakefield literature~\cite{Sprangle_PRA_1990, Mora_PoP_1997, Huang_JCP_2006, Diederichs_CPC_2022, Lindstrom_arXiv_2025}, we extend it here to capture streaming instabilities at all spatiotemporal scales.

Combining the linearized plasma electron fluid equations with Maxwell’s equations for the scalar ($\phi$) and vector ($\mathbf{A}$) potentials yields under the QSA~\cite{San_Miguel_Claveria_long_paper_2026} 
\begin{align}
    &(\partial_\xi^2 + k_p^2) n_p^{(1)} = k_p^2 \frac{\rho_b^{(1)}}{e} \,, \label{eq:plasmaeq} \\
    &(\partial_y^2-k_p^2) (A_x^{(1)}-\phi^{(1)}/c) = -\frac{e n_p^{(1)}}{\epsilon_0 c}  \,, \label{eq:Poissoneq}
\end{align}
where $\rho_b = \sum_s Z_s e n_{bs}$ is the total beam charge density. These equations are generally used to describe the plasma wakefield driven by a ballistic particle beam \cite{Chen_IEEE_1987, Keinigs_POF_1987}; here, they describe the coupling of beam, plasma and field perturbations. The pseudo-potential $\Psi \equiv A_x-\phi/c$ encapsulates both the electrostatic and electromagnetic fields of the excited plasma waves~\cite{Mora_PoP_1997, Lindstrom_arXiv_2025}.

To close the system, we evolve the beam over $\tau$. Combining the linearized continuity and momentum equations for each beam species $s$, we find~\cite{San_Miguel_Claveria_long_paper_2026}
\begin{equation}
    \partial_\tau^2 n_{bs}^{(1)} + \frac{Z_s e c}{\gamma_b M_s m_e} n_{bs}^{(0)} \left(\gamma_b^{-2}\partial_\xi^2 + \partial_y^2 \right) \Psi^{(1)} = 0 \,.
\label{eq:beameq}
\end{equation}
Gathering Eqs.~\eqref{eq:plasmaeq}, \eqref{eq:Poissoneq} and \eqref{eq:beameq}, we derive the equation that governs the perturbation:
\begin{align}   
    &\left[ (\partial_\xi^2+k_p^2)(\partial_y^2-k_p^2)\partial_\tau^2 -
    k_p^2 \sum_s \omega_{bs}^2\left(\frac{\partial_\xi^2}{\gamma_b^2}+\partial_y^2 \right) \right] \Psi^{(1)} \nonumber \\
    &= 0 \,.
    \label{eq:master_QSA}
\end{align}
This equation describes the spatiotemporal evolution of the full 2D electromagnetic spectrum of unstable modes and, crucially, holds beyond the SVEA.

In the relativistic cold-beam limit, the dominant modes have a finite transverse wavenumber $k_y$ \cite{Bludman_POF_1960, Bret_POP_2010}. Expressing $\Psi^{(1)} = \delta \Psi(\xi,\tau) \exp \left(ik_y y\right)$ and neglecting $\gamma_b^{-2}\partial_\xi^2 \ll k_y^2$, we obtain
\begin{equation}
    \left[(\partial_\xi^2+k_p^2)\partial_\tau^2 - k_p^2 \Gamma^2 \right] \delta \Psi = 0 \,,
    \label{eq:mastereq_ST_CFI}
\end{equation}
where $\Gamma = \left(\sum_s \omega_{b s}^2\frac{k_y^2}{k_y^2+k_p^2}\right)^{1/2}$ is the purely temporal CFI growth rate in unbounded geometry~\cite{Bret_POP_2010}. Further writing $\delta \Psi = \widehat{\delta \Psi}(\xi,\tau) \exp (ik_p \xi)$, and applying the SVEA $(\partial_\xi\ll k_p)$, we recover the spatiotemporal equation of the electrostatic OTSI under the QSA ~\cite{San_Miguel_Claveria_PRR_2022, Walter_PRE_2024}.

\begin{figure}
    \includegraphics[width=0.49\textwidth]{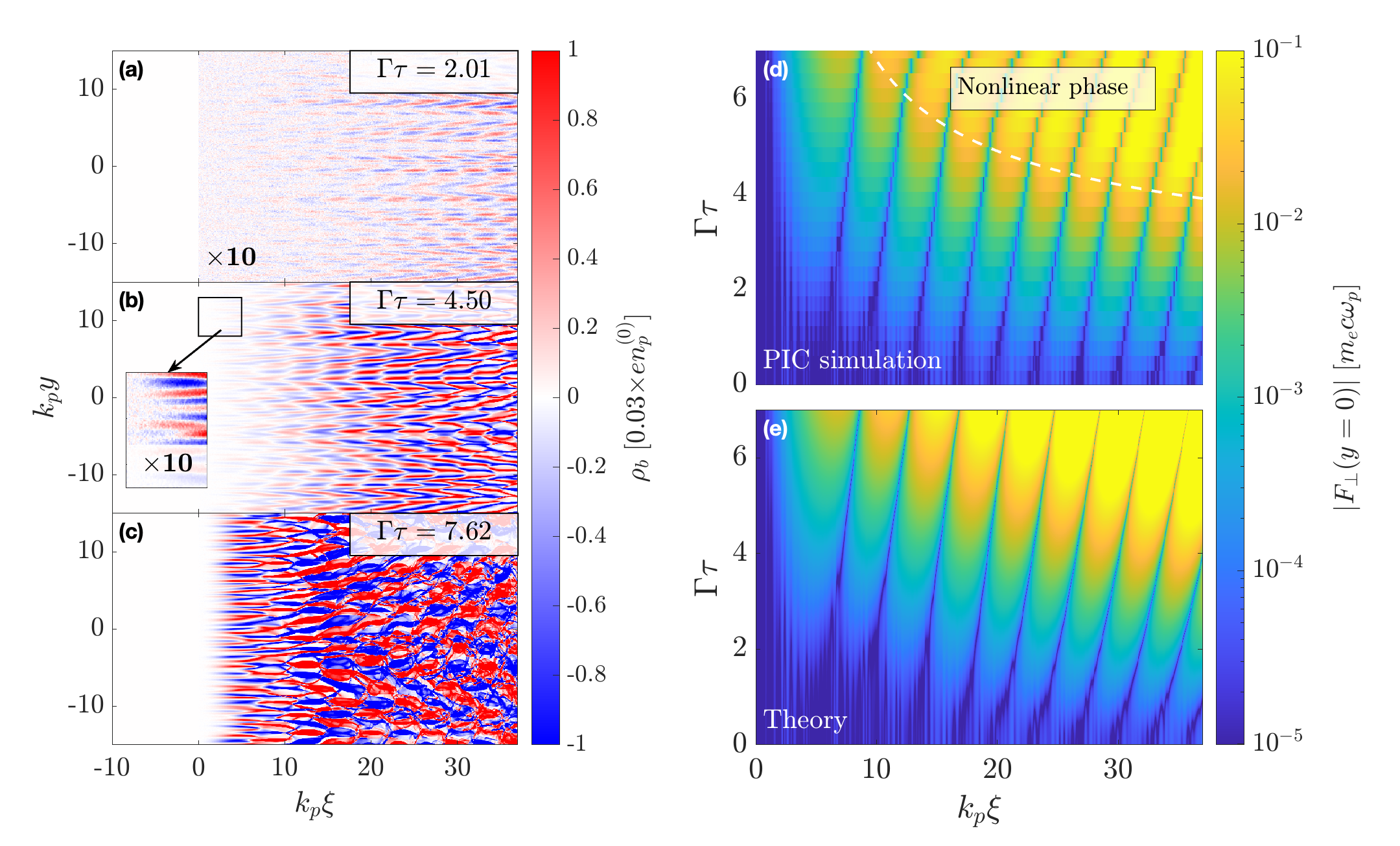}
    \caption{2D PIC pair beam-plasma simulation with $\alpha_b = 0.03$ and $\gamma_b = 2\times 10^4$ for both beam electrons and positrons. (a–c) Snapshots of the beam charge density $\rho_b(\xi,y)$ at three successive times. In (b), the inset zooms in on the beam-front region delineated by the small box. (d, e) $\xi$--$\tau$ evolution of $\vert F_\perp(y=0) \vert$ as (d) extracted from the PIC simulation or (e) computed from Eq.~\eqref{eq:mastereq_ST_CFI} (see text for details). In (d), the dashed line marks the transition to the nonlinear instability regime.
    \label{fig:1}}
\end{figure}

\textit{PIC simulations.---}
To validate Eq.~\eqref{eq:mastereq_ST_CFI}, we compare its numerical solution to a \textsc{calder}~\cite{Lefebvre_NF_2003} PIC simulation of a relativistic electron-positron beam injected into an electron-ion plasma. The simulation is performed in a 2D3V geometry (two-dimensional in space, three-dimensional in momentum space) with an advanced numerical scheme to suppress numerical Cherenkov-like instabilities~\cite{Lemoine_PRL_2019, Vanthieghem_AJL_2022}. The beam propagation is followed by a moving window, extending over $-10 \le k_p\xi \le 40$ and $-20 \le k_p y \le 20$ with mesh size $k_p\Delta x = k_p\Delta y = 0.05$. The time step is $\omega_p\Delta t = 0.045$. The beam density profile is flat-top over $0 \le k_p\xi \le 37$, with $\alpha_b = 0.03$ and $\gamma_b = 2\times 10^4$ for both electrons and positrons (smoother beam profiles with linear or $\sin^2$ ramps over $1$--$10\,k_p^{-1}$ yield very similar results). Plasma ions are kept immobile, while other species are initialized at zero temperature with ten macroparticles per cell. Boundary conditions are open along $\xi$ and periodic along $y$ for both particles and fields. Instabilities develop spontaneously from numerical noise.

Figures~\ref{fig:1}(a-c) show the beam charge density at three successive times. Chevron-type modulations, indicative of OTSI, develop throughout the beam, except near its front (within a few $k_p^{-1}$), where they evolve into longitudinal filaments characteristic of a locally dominant CFI. Figure~\ref{fig:1}(d) displays the $\xi$--$\tau$ evolution of the transverse Lorentz force $F_\perp \equiv e(E_y - cB_z)$ (normalized to $m_e \omega_p c$) measured along $y = 0$. Since $F_\perp \propto \nabla_\perp \Psi$, its evolution is expected to be dictated by Eq.~\eqref{eq:mastereq_ST_CFI}.

The numerical solution to this equation, using $\Gamma = 1.7 \times 10^{-3}\,\omega_p$ as computed from simulation data ($k_y/k_p \simeq 5.5$), is plotted in Fig.~\ref{fig:1}(e). The initial profile $F_\perp(\xi,\tau=0)$ is taken from the simulation and the boundary value is set to $F_\perp(\xi=0,\tau>0) = F_\perp(\xi=0,\tau=0)$. The agreement between theory and simulation is excellent in both amplitude and phase across the simulation window. Discrepancies appear only upon transition to the nonlinear regime [above the dashed line in Fig.~\ref{fig:1}(d)], which is outside the scope of our model. Note that Eq.~\eqref{eq:mastereq_ST_CFI} generalizes to arbitrary longitudinal beam profiles via $\Gamma(\xi) = \left(\sum_s \omega^2_{b s}(\xi) \frac{k_y^2}{k_y^2+k_p^2}\right)^{1/2}$, thus making it applicable to realistic high-energy accelerator beams regardless of their length relative to $k_p^{-1}$ \cite{San_Miguel_Claveria_long_paper_2026}.

\textit{Analytical solutions.---}
The simulation results can be understood from the analytical properties of the solution to Eq.~\eqref{eq:mastereq_ST_CFI} for an initial pulse disturbance at $\xi=\tau=0$. After applying a double Laplace transform~\cite{Decker_POP_1996}, a saddle-point analysis reveals two distinct instability regimes in the time-asymptotic limit ($\Gamma \tau > 1$) \cite{San_Miguel_Claveria_long_paper_2026}.

Far enough from the beam front, $k_p\xi \gg 2\Gamma \tau/\sqrt{27}$, the amplitude of the dominant mode evolves as
\begin{align}
    \delta \Psi (\tau,\xi) \propto \exp\left( i k_p\xi + \frac{3\sqrt{3}}{4}(\Gamma \tau)^{2/3} (k_p\xi)^{1/3}\right) 
     \,.
    \label{eq:beam_rear}
\end{align}
Given the temporal OTSI growth rate $\Gamma_{\rm OTSI}=\frac{\sqrt{3}}{2^{4/3}} \Gamma^{2/3}$ for unbounded systems \cite{Bludman_POF_1960, Bret_POP_2010}, this solution represents the spatiotemporal variant of OTSI previously treated in the purely electrostatic limit~\cite{San_Miguel_Claveria_PRR_2022}.

Conversely, in the near-front region ($k_p\xi \ll 2\Gamma \tau/\sqrt{27}$), the dominant mode grows as 
\begin{align}
    \delta \Psi(\tau,\xi) \propto
    \exp\left(2 \sqrt{\Gamma \tau k_p \xi}\,(1-k_p\xi/4\Gamma \tau)\right) \,.
    \label{eq:beam_front}
\end{align}
The non-oscillatory behavior of this solution distinguishes it as spatiotemporal CFI. In fact, a purely growing dynamics persists up to $k_p\xi = 2 \Gamma \tau/\sqrt{27}$, where an observer would experience an effective growth rate of $(4/3)\sqrt{2/3} \Gamma \simeq 1.09 \Gamma$ \cite{San_Miguel_Claveria_long_paper_2026}. Our unified framework---incorporating electrostatic and electromagnetic effects through the pseudo-potential---remains valid where the SVEA fails.

\begin{figure}
    \includegraphics[width=0.49\textwidth]{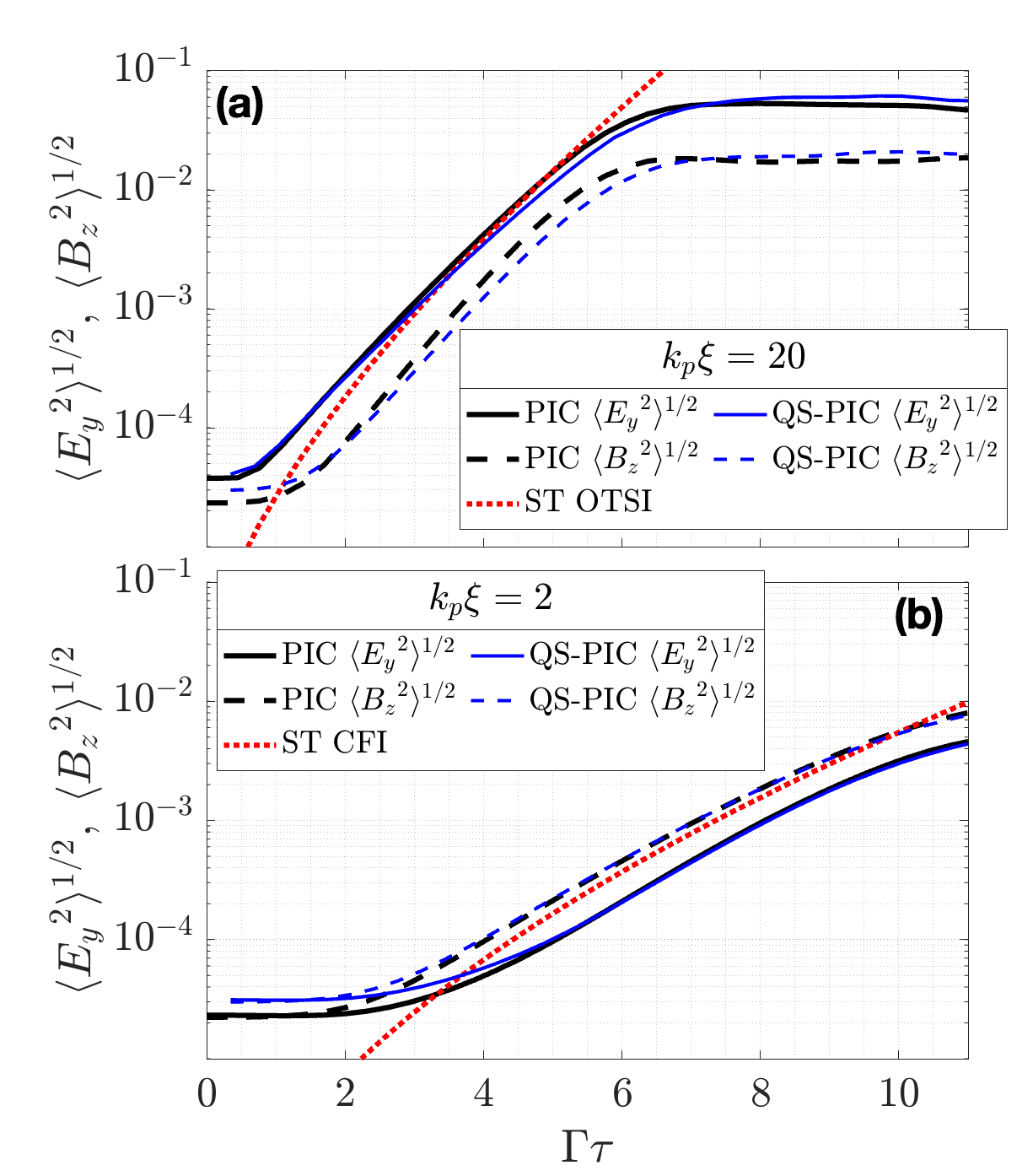}
    \caption{Transverse electric ($\langle E_y^2\rangle^{1/2}$, solid lines) and magnetic ($\langle B_z^2\rangle^{1/2}$, dashed lines) field amplitudes as a function of the normalized propagation time ($\Gamma \tau$), at two locations from the beam front: (a) $k_p\xi = 20$ and (b) $k_p\xi = 2$. Black lines correspond to the PIC simulation presented in Fig.~\ref{fig:1}, while blue lines represent an equivalent QS-PIC simulation. The scalings of spatiotemporal (ST) OTSI [Eq.~\eqref{eq:beam_rear}] and CFI [Eq.~\eqref{eq:beam_front}] are plotted as red solid and dashed lines in (a) and (b), respectively. 
    \label{fig:2}}
\end{figure}

\begin{figure*}[ht!]
    \centering 
    \includegraphics[width=\textwidth]{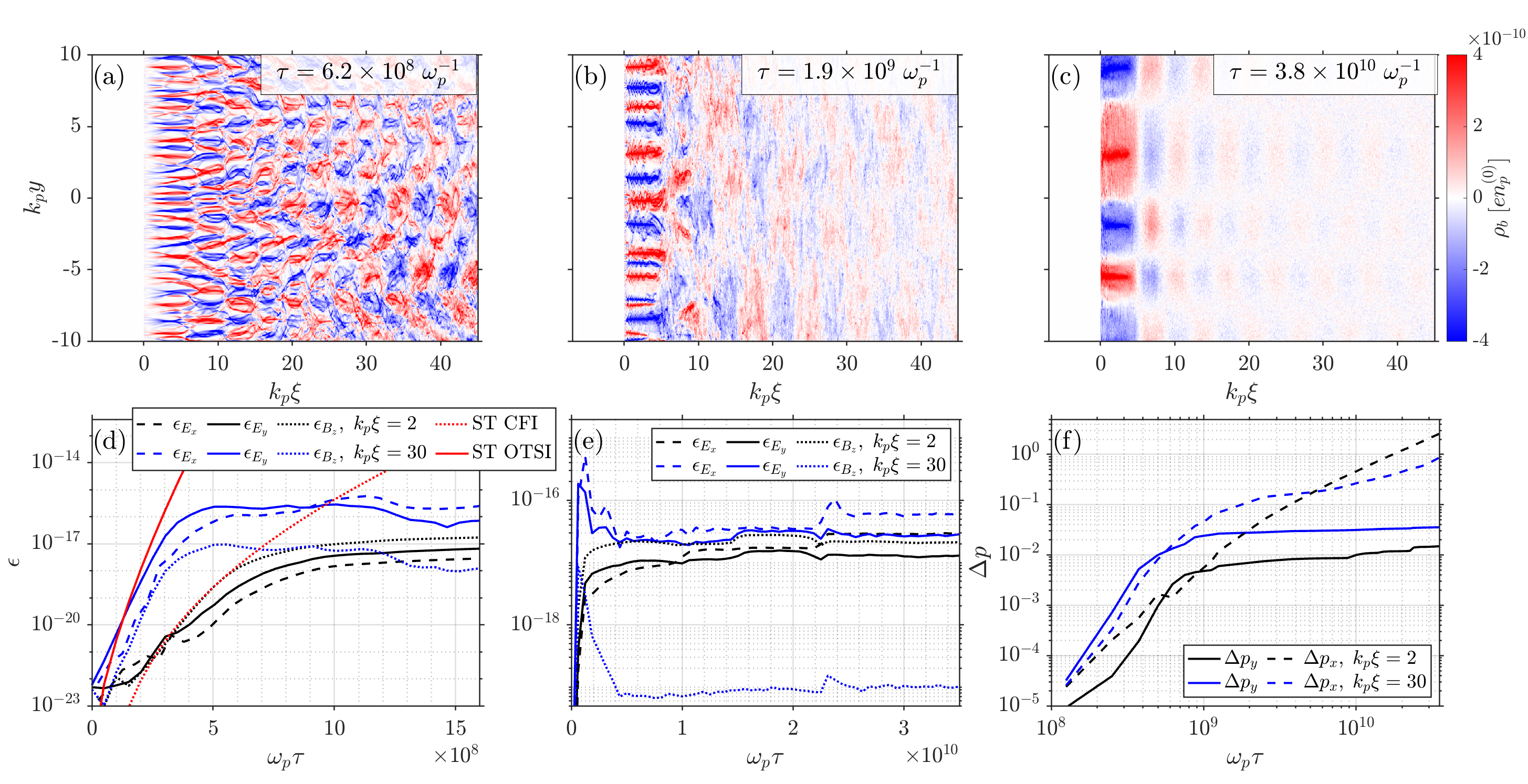}
    \caption{2D QS-PIC pair beam-plasma simulation with parameters relevant for blazar jets ($\alpha_b=10^{-10}$ and $\gamma_b=10^6$ for both beam electrons and positrons). (a-c) Beam charge density profile (normalized to $en_p^{(0)}$) at three successive times. (d) Early-time evolution of the $E_x$ (solid), $E_y$ (dashed) and $B_z$ (dotted) field energies ($\epsilon$, normalized to the initial beam kinetic energy) at two positions along the beam: $k_p\xi = 2$ (black) and $k_p\xi = 30$ (red). (e) Same as (d) but for the full simulation duration. (f) Evolution of the root-mean-square momentum spread of beam particles along the longitudinal (dashed) and transverse (solid) directions at the same positions as in (d-e).
    \label{fig:fig3}}
\end{figure*}

Figure~\ref{fig:2} compares these analytical solutions (red lines) to PIC results (black lines), displaying root-mean-squared amplitudes of the transverse electric ($\langle E_y^2 \rangle^{1/2}$) and magnetic ($\langle B_z^2 \rangle^{1/2}$) fields, averaged over the transverse domain and longitudinal segments of width $\Delta \xi = 2\,k_p^{-1}$. At $k_p\xi= 20$ [Fig.~\ref{fig:2}(a)], field growth is dominated by $E_y$ and follows the prediction for spatiotemporal (ST) OTSI [Eq.~\eqref{eq:beam_rear}] until saturation sets in at $\Gamma \tau \simeq 6$. By contrast, near the front at $k_p\xi= 2$ [Fig.~\ref{fig:2}(b)], $B_z$ dominates, aligning with the spatiotemporal CFI scaling of Eq.~\eqref{eq:beam_front} throughout the simulation duration.

This agreement confirms the predicted spatiotemporal hierarchy where, in the linear regime, CFI prevails at the beam front and OTSI dominates downstream. Although demonstrated analytically here for a pulse disturbance at the beam front, this behavior also holds for an instability seed extending throughout the beam~\cite{San_Miguel_Claveria_long_paper_2026}. Interestingly, the transition region scaling ($k_p \xi \lesssim 2 \Gamma \tau/\sqrt{27}$) suggests that after a given number of temporal $e$-folds $N_e$ (as applies at the transition point), the CFI modes extend over a distance $\sim 0.4 N_e k_p^{-1}$ from the front, independent of the beam density ratio and Lorentz factor. This prediction is consistent with Fig.~\ref{fig:1}(c), which shows that at saturation ($N_e \sim 10$), filaments are confined to $k_p \xi \lesssim 5$. We have further verified it through additional simulations across a wide range of $\alpha_b \in [10^{-10}, 10^{-2}]$ and $\gamma_b \in [10^2, 10^6]$. These results invalidate previous findings from a purely inductive, SVEA-based model~\cite{Pathak_NJP_2015} that spatiotemporal CFI is confined to $\xi/c\tau \propto \alpha_b/\gamma_b^3$ and is hence inoperative at very large $\gamma_b$ and small $\alpha_b$.

\textit{Extending the QSA to extreme blazar jet regimes.---}
While QS-PIC codes have become standard for plasma wakefield accelerators~\cite{Huang_JCP_2006, Mehrling_PPCF_2014, Sosedkin_NIMA_2016, Diederichs_CPC_2022}, their ability to model ultrarelativistic streaming instabilities in extreme astrophysical regimes remains largely unexplored. We demonstrate this capability using \textsc{QuaSSis}, a new 2D3V QS-PIC code implementing transversely periodic boundary conditions and random macroparticle weights for precise noise control~\cite{Labro_CPC_2026}. Benchmarking against the \textsc{calder} run of Fig.~\ref{fig:2} shows identical field evolution but with considerable speedup: \textsc{QuaSSis} completed the simulation in $\sim 1$~h on a single core, compared to $\sim 3\times10^3$ core-hours for full PIC. This efficiency gain, scaling as $\sim \omega_p/\omega_b$, unlocks previously inaccessible parameter spaces.

A particularly challenging problem is that of blazar jets~\cite{Broderick_APJ_2018}, where TeV $\gamma$-rays decay into pair beams propagating with $\gamma_b \sim 10^6$ through the much denser intergalactic medium ($\alpha_b \sim 10^{-15}$--$10^{-10}$). The corresponding dynamical range ($\omega_p/\omega_b \sim 10^8$--$10^{10}$) is clearly inaccessible to conventional PIC simulations, which have been limited so far to $\omega_p/\omega_b \lesssim 1000$ ($\alpha_b \gtrsim 10^{-4}$, $\gamma_b \lesssim 10^3$) \cite{Sironi_APJ_2014, Vafin_APJ_2018, Peterson_APJL_2022}. To showcase its potential, we push \textsc{QuaSSis} to an unprecedented regime with $\alpha_b = 10^{-10}$ and $\gamma_b = 10^6$, capturing the transition from linear growth to deep nonlinearity over several $10^{10}\omega_p^{-1}$. 

At early times [Figs.~\ref{fig:fig3}(a,d)], the expected hierarchy emerges: spatiotemporal CFI dominates the beam front ($k_p\xi \lesssim 5$), while spatiotemporal OTSI prevails further downstream. Their respective dynamics match the analytical predictions until reaching the saturation regime, which occurs at $\omega_p \tau_\mathrm{sat} \simeq 7\times 10^8$ for CFI ($k_p\xi = 2$) and $\omega_p \tau_\mathrm{sat} \simeq 5\times 10^8$ for OTSI ($k_p\xi = 30$). Saturation is attributed to beam trapping, which arises when the transverse bounce frequency $\omega_B$ of the beam particles in the locally dominant field modulation approaches the \textit{spatiotemporal} growth rate of the corresponding instability, $\Gamma^{\rm st} = \partial_\tau \ln \vert \Psi(\tau,\xi) \vert$---rather than the temporal growth rate usually considered for this criterion \cite{Davidson_POF_1972, Bresci_PRE_2022}. In turn, this leads to a transverse beam momentum spread increasing to $\Delta p_y/m_e c \simeq \gamma_b \omega_B/k_y c \simeq \gamma_b \Gamma^{\rm st} /k_y c$. For CFI, one has $\omega_B = \sqrt{ek_y cB_z/m_e \gamma_b}$ and $\Gamma^{\rm st}/\omega_p = \sqrt{(\Gamma/\omega_p) \xi/c\tau}$. For $k_p \xi = 2$ and $k_y \simeq k_p$, we thus predict a fractional magnetic field energy $\epsilon_{B_z} = B_z^2/(4\mu_0 n_b \gamma_b m_e  c^2) \simeq (\omega_p/k_y c)^2 (\xi/c\tau_\mathrm{sat})^2/4 \simeq 2\times 10^{-18}$ at saturation, associated with $\Delta p_y/m_e c \simeq 5\times 10^{-3}$. For OTSI, one has $\omega_B = \sqrt{ek_y c E_y/m_e \gamma_b}$ and $\Gamma^{\rm st}/\omega_p = (\sqrt{3}/2) (\Gamma/\omega_p)^{2/3}(\xi/c\tau)^{1/3}$, giving a saturated fractional electric field energy $\epsilon_{E_y} = \epsilon_0 E_y^2/(4 n_b \gamma_b m_e c^2) \simeq (9/64 k_y^2) (n_b/\gamma_b)^{1/3} (\xi/c\tau_\mathrm{sat})^{4/3} \simeq 1.5 \times 10^{-16}$ and a momentum spread $\Delta p_y/m_e c \simeq 0.016$. These estimates agree with simulation data [Figs.~\ref{fig:fig3}(d,f)]. Downstream, OTSI proceeds nonlinearly such that the electric field energy becomes dominated by $E_x$ \cite{Sironi_APJ_2014}, which reaches its peak at $\omega_p \tau \simeq 10^9$ before dropping by an order of magnitude around $\omega_p \tau \simeq 4\times 10^9$ [Fig.~\ref{fig:fig3}(e)].

Interestingly, at this stage, the beam front has fragmented into pinched current filaments (with length $l \sim 5\,k_p^{-1}$, width $w \sim 0.5$--$1\,k_p^{-1}$ and spacing $\lambda \sim 2\,k_p^{-1}$), driving a spectrum of plasma wakefields with $k_y \simeq \pi/\lambda \sim k_x$. The downstream particles are too hot transversely to further amplify these waves, so their amplitude is determined primarily by the filaments as $\vert eE_x/m_e c \omega_p\vert \simeq \vert 2 (\rho_b/en_p^{(0)}) \sin (k_p l/2) \vert (1-e^{-k_p w})$ \cite{Chen_IEEE_1987}. This amplitude increases slightly as the filaments widen via coalescence \cite{Medvedev_APJ_2005} to $k_p w \sim 2$--3 [Fig.~\ref{fig:fig3}(c)]. One such event at $\omega_p \tau \simeq 2.4\times 10^{10}$ explains the approximate doubling in downstream $E_x$ energy seen in Fig.~\ref{fig:fig3}(e). Interaction with these wakefields progressively broadens the longitudinal beam momentum spread as $\Delta p_x/m_e c \sim \vert \rho_b/e n_p^{(0)} \vert \omega_p \tau$, consistent with Fig.~\ref{fig:fig3}(f). 

The robust agreement between the simulation and the above nonlinear scalings suggests that QS-PIC codes can accurately describe the nonlinear dynamics of ultrarelativistic streaming plasma instabilities. This opens the way to exciting physics studies at longer timescales, where quasi-parallel two-stream instability or temporal CFI in the heated beam bulk~\cite{Sironi_APJ_2014} could emerge and alter the scaling $\Delta p_x \propto \tau$. This also enables higher-fidelity studies incorporating larger initial beam momentum spreads, ion motion~\cite{Chang_APJ_2014, Vafin_APJ_2019}, external magnetic fields~\cite{Alawashra_APJ_2022}, or plasma inhomogeneities~\cite{Perry_MNRAS_2021}, which should shine new light on the plasma dynamics involved in blazar jet regimes and will be the subject of future work.

\textit{Conclusion.---}
We have developed a general spatiotemporal theory of streaming plasma instabilities based on the QSA. By avoiding the SVEA, our model captures the full unstable spectrum, including electrostatic and electromagnetic effects. For a bounded, ultrarelativistic cold beam, we unravel the competition between OTSI and CFI, showing that magnetic CFI modes are confined to the beam front. The agreement between theory, full PIC, and QS-PIC simulations underscores the power of the QSA to reduce computational costs by orders of magnitude. This framework opens a new avenue to exploring ultrarelativistic streaming plasma instabilities. It is particularly relevant to laboratory beams where finite beam size imposes strong spatiotemporal coupling, but also to astrophysical scenarios like blazar jets, suggesting a previously unconsidered interplay between OTSI and CFI---namely, that the post-OTSI bulk beam dynamics is dominated by wakefields from front-side, CFI-driven current filaments. Our results offer new insights into the instabilities involved in blazar jet regimes and the capacity to efficiently capture them numerically. A companion paper extends this theory to purely longitudinal modes, including the two-stream, self-modulation, and hosing instabilities~\cite{San_Miguel_Claveria_long_paper_2026}.

\begin{acknowledgments}
P.S.M.C. received funding from the European Union under HORIZON-WIDERA-2023-TALENTS-02 (PLAXI project, Grant Agreement No.~101180632). The work at IST was supported by the European Research Council (ERC-2021-CoG Grant XPACE No.~101045172). The work at CEA and LOA was supported by the Agence Nationale de la Recherche (ANR) (UnRIP project, Grant No.~ANR-20-CE30-0030). Computational resources were provided by FCT I.P. on the Deucalion and MareNostrum5 platforms through Project 2025.00196.CPCA.A3 and 2025.00204.CPCA.A3, and by GENCI-TGCC on the supercomputer IRENE under Grants No.~A0100510786 and No.~A0190512993.

\end{acknowledgments}

\bibliography{biblio}

\end{document}